\documentclass[twocolumn,showpacs,preprintnumbers,amsmath,amssymb]{revtex4}

\usepackage{graphicx}
\usepackage{dcolumn}
\usepackage{bm}
\usepackage{amsmath}

\begin{document}
\title{Greatly enhancing the modeling accuracy for distributed parameter systems by nonlinear time/space separation}
\author{Hai-Tao Zhang$^{1}$}
\author{Chen-Kun Qi$^{2}$}
\author{Tao Zhou$^{3}$}
\email{zhutou@ustc.edu}
\author{Han-Xiong Li$^{2}$}

\affiliation{%
$^{1}$Department of Control Science and Engineering, Huazhong
University of Science and Technology, Wuhan 430074, PR China \\
$^{2}$Department of Manufactory Engineering and Engineering
Management, City University of Hong Kong, Hong Kong SAR, PR China \\
$^{3}$Department of Modern Physics and Nonlinear Science Center,
University of Science and Technology of China, Hefei 230026, PR
China
}%

\date{\today}

\begin{abstract}
An effective modeling method for nonlinear distributed parameter
systems (DPSs) is critical for both physical system analysis and
industrial engineering. In this paper, we propose a novel DPS
modeling approach, in which a high-order nonlinear Volterra series
is used to separate the time/space variables. With almost no
additional computational complexity, the modeling accuracy is
improved more than 20 times in average comparing with the
traditional method.
\end{abstract}

\pacs{05.45.Gg, 02.30.Yy, 07.05.Dz}

\maketitle

\emph{Introduction} - Most of the physical processes (e.g. thermal
diffusion process
\cite{Christofides2001,Hoffman1998,Chechkin2002,Sheintuch2004,Kuptsov2005,Kossler2006,Yagi2004},
thermal radiation process \cite{Kazakov1994}, distributed quantum
systems \cite{Katsnelson2000,Wakabayashi1998}, concentration
distribution process \cite{Vlad2002,Hass1995,McGraw2003}, crystal
growth process \cite{Christofides2001,Kossler2006}, etc.) are
nonlinear distributed parameter systems (DPSs) with boundary
conditions determined by the system structure. Thus, it is an
urgent task to design an effective modeling method for nonlinear
DPSs. The key problem in the design of nonlinear-DSP modeling
method is how to separate the time/space variables. Some modeling
approaches are previously proposed: These include the
Karhunen-Lo\`{e}ve (KL) approach
\cite{Christofides2001,Sheintuch2004,Ray1981,Park1996}, the
spectrum analysis \cite{Gottlieb1993}, the singular value
decomposition (SVD) combined with the Galerkin's method
\cite{Christofides2001,Chakravarti1995}, and so on. Among them,
the KL approach is the most extensively studied and the most
widely applied one. In this approach, the output $T(z,t)$ is
expanded as
\begin{equation}
T(z,t)=\sum^N_{i=1}c_i(z)l_i(t)\triangleq C(z)L(t),
\end{equation}
where $z$ and $t$ are the space and time variables, respectively.
This operation can be implemented by spatial basis $\{c_i(z)\}$
combined with time-domain coefficients $\{l_i(t)\}$, or
time-domain basis $\{l_i(t)\}$ combined with spatial coefficients
$\{c_i(z)\}$. The basis could be Jacobi series \cite{Datta1995},
orthonormal functional series (OFS, such as Laguerre series
\cite{Datta1995,Hu2004}, Kautz series \cite{Datta1995}, etc.),
spline functional series \cite{Boor1978}, trigonometric functional
series, or some others. However, Nno matter how elaborately the
basis is designed, the infinite-dimensional nature of DPSs does
not allow being accurately modeled with small number of truncation
length $N$ of the basis series. Moreover, the nonlinear nature of
the DPSs will even increase this modeling difficulty. Thus for
nonlinear DPSs' modeling, the extending of $N$ to a sufficiently
large number is generally required, which would definitely
increase the computational burden. Consequently, in order to
improve the efficiency of the modeling algorithms, many former
efforts focused on designing suitable time-domain basis $\{
l_i(t)\}$ or spatial basis $\{ c_i(z)\}$ according to the prior
knowledge of the DPSs \cite{Christofides2001,Ray1981}. In
addition, some scholars also presented neural networks to model
the nonlinearities of transitional flows and distributed reacting
systems based on proper orthogonal decomposition and Galerkin's
method \cite{Sahan1997,Shvartsman2000}. However, if the prior
knowledge is unavailable or inadequate, the general design methods
of the bases are very limited so far. On the other hand, the
conventional finite difference and finite element method often
lead to very high-order ODEs which are inappropriate for dynamical
analysis and real-time implementation \cite{Shvartsman2000}.
Another conventional approach, spectral method
\cite{Gottlieb1993}, is popularly used for modeling DPSs because
it may result in very low dimensional ODE systems. However, it
requires a regular boundary condition
\cite{Christofides2001,Canuto1988}.

Thus, in this paper, we argue that the linear separation is a
bottleneck to better modeling performance, and to introduce some
kinds of nonlinear terms may sharply enhance the performance,
since they have the capability to compensate the residuals of the
linear separation.

\emph{The Implement of Nonlinear Space/Time Separation} - For
nonlinear lumping systems, if their dependencies on past inputs
decrease rapidly enough with time, their input/output relationship
can be precisely described by \textbf{Volterra series}
\cite{Schetzen1980,Boyd1985,Kuz1991,Cherdantsev1995,Chen1999},
which is a generalization of the convolution description of linear
time-invariant to time-invariant nonlinear operators. This kind of
system is called fading memory nonlinear system (FMNS)
\cite{Boyd1985}, which is well-behaved in the sense that it will
not exhibit multiple steady-states or other related phenomena like
chaotic responses. Fortunately, most industrial processes are
FMNSs. Accordingly, one can naturally extend the concept of
Volterra series from lumping systems to DPSs by allowing each
kernel to contain both time variable $t$ and space variable $z$,
and then design the time/space separation method via the so-call
\textbf{distributed Volterra series} (see Fig. 1 for the mechanism
of this modeling method). Firstly, the system output can be
represented by
\begin{eqnarray}
T(z,t)=h_0(z)+\int^{\infty}_0h_1(z,\tau_1)u(t-\tau_1)d\tau_1+ \nonumber\\
\int^{\infty}_0\int^{\infty}_0h_2(z,\tau_1,\tau_2)u(t-\tau_1)u(t-\tau_2)d\tau_1d\tau_2+\cdots,
\end{eqnarray}
where $h_i(z,\tau_1,\cdots,\tau_i)$ is the $i$th order distributed
Volterra kernel. Then denote $\phi_i(t)$ as the $i$th order OFS
and $l_i(t)=\int^{\infty}_0\phi_i(t)u(t-\tau)d\tau$ as the $i$th
order OFS filter output. Since $\{\phi_i(t)\}$ forms a complete
orthonormal set in functional space $l_2$, each kernel can be
approximately represented as
\begin{eqnarray}
h_1(z,\tau_1)=\sum^N_{k=1}c_k(z)\phi_k(\tau_1), \nonumber\\
h_2(z,\tau_1,\tau_2)=\sum^N_{n=1}\sum^N_{m=1}c_{nm}(z)\phi_n(\tau_1)\phi_m(\tau_2), \\
\cdots \nonumber
\end{eqnarray}
where $c_k(z)$ and $c_{nm}(z)$ are spatial coefficients. Then, the
input/output relationship can be written as (see Eq. (1) for
comparison)
\begin{equation}
T(z,t)=c_0(z)+C(z)L(t)+L^T(t)D(z)L(t)+\cdots,
\end{equation}
where $L(t)=[l_1(t) \cdots l_N(t)]^T$, $C(z)=[c_1(z) \cdots
c_N(z)]$, and $D(z)=[c_{ij}(z)]_{N\times N}$.

To obtain the spatial coefficients, firstly we pre-compute all the
OFS kernels according to the polynomial iterations
\cite{Datta1995} or the following state equation
\begin{equation}
L(t+1)=AL(t)+Bu(t),
\end{equation}
where $u(t)$ is the system input, and $A$ and $B$ are
pre-optimized matrices (see Ref. \cite{Wang2004} for details).
Then the input/output relationship Eq. (4) is represented by a
linear regressive form, and these spatial coefficients $c_0(z)$,
$C(z)$, $D(z)$, $\cdots$ can be obtained by using the least square
estimation combined with spline interpolation \cite{Boor1978}.
Finally, the model is obtained by synthesizing the OFS kernels and
the spatial coefficients according to Eq. (4).

\begin{figure}
\scalebox{0.3}[0.25]{\rotatebox{90} {\includegraphics{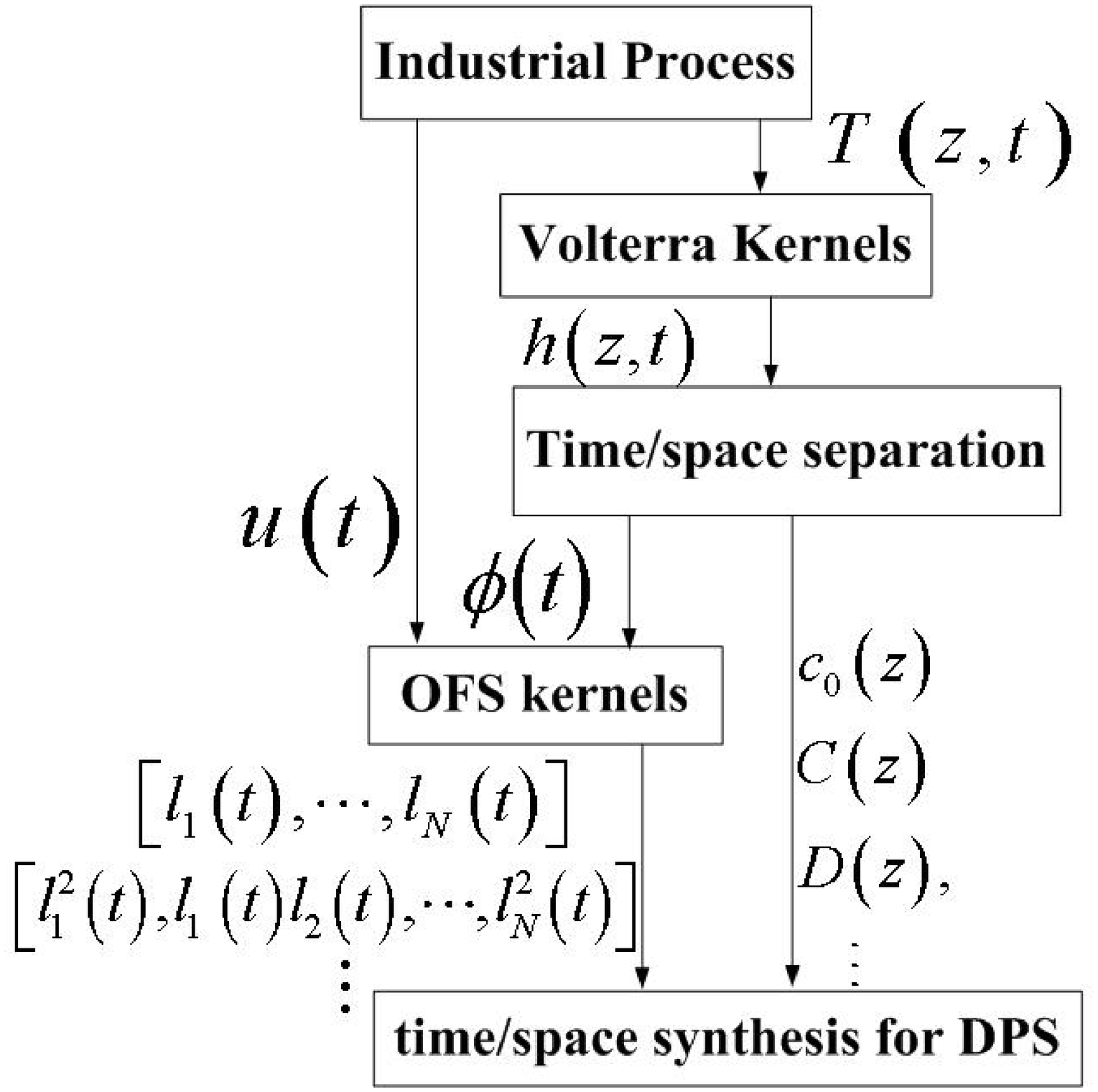}}}
\caption{The sketch map of OFS-Volterra modeling for nonlinear
DPS.}
\end{figure}
\begin{figure}
\scalebox{0.3}[0.3]{\rotatebox{90} {\includegraphics{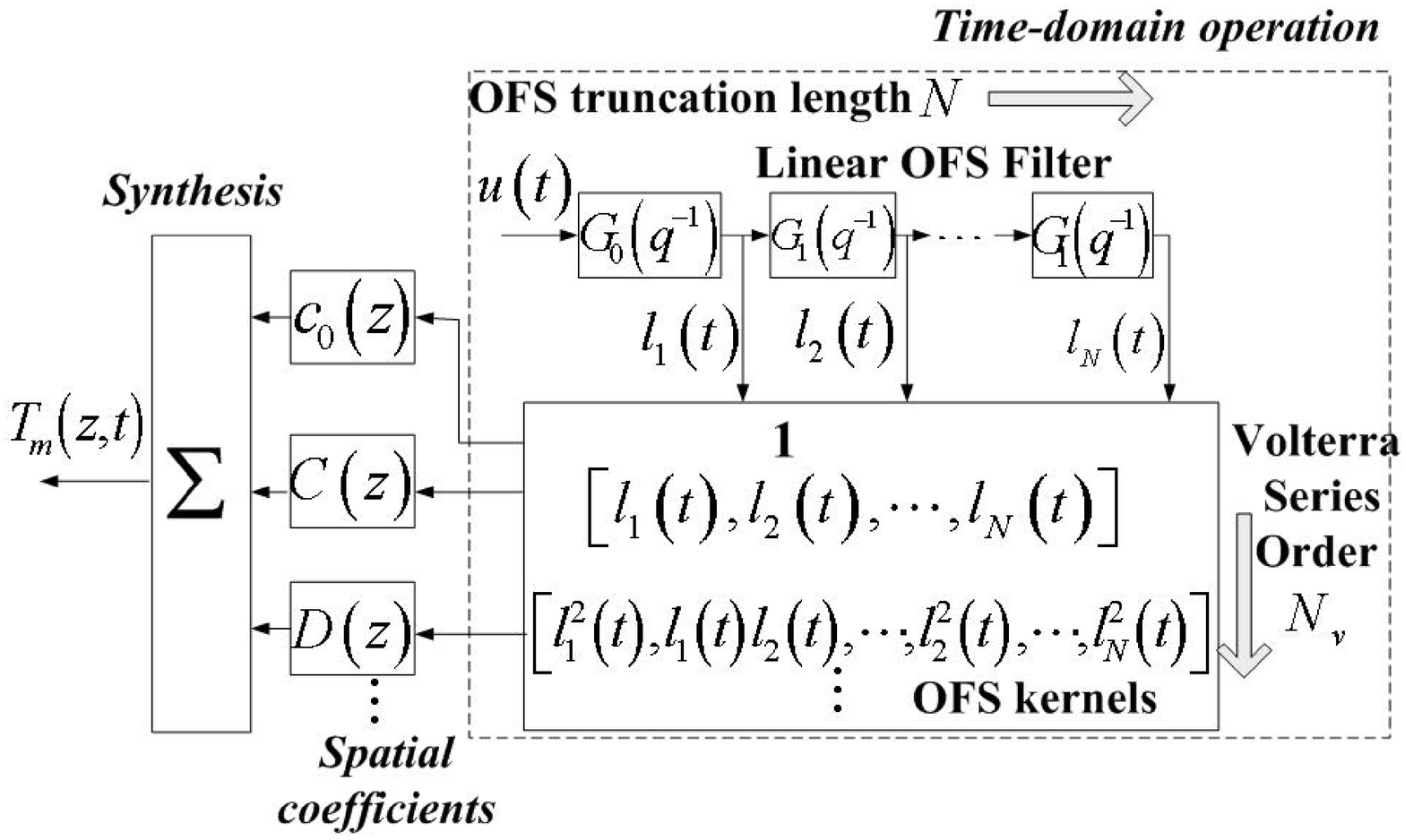}}}
\caption{Operation details of OFS-Volterra modeling.}
\end{figure}
\begin{figure}
\scalebox{0.20}[0.2]{\rotatebox{90} {\includegraphics{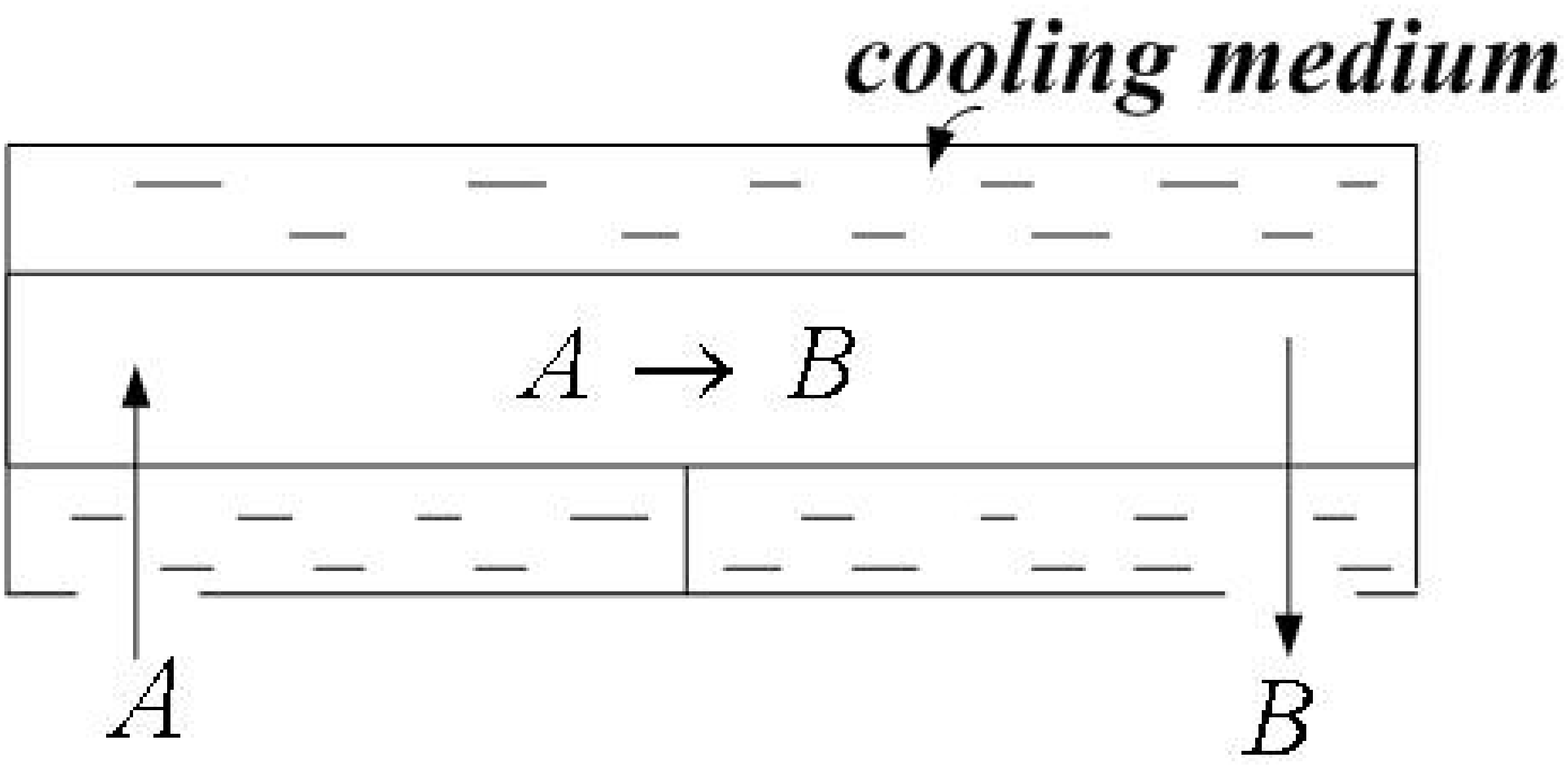}}}
\caption{The sketch map of catalytic rod.}
\end{figure}
\begin{figure}
\scalebox{0.25}[0.25]{\rotatebox{90} {\includegraphics{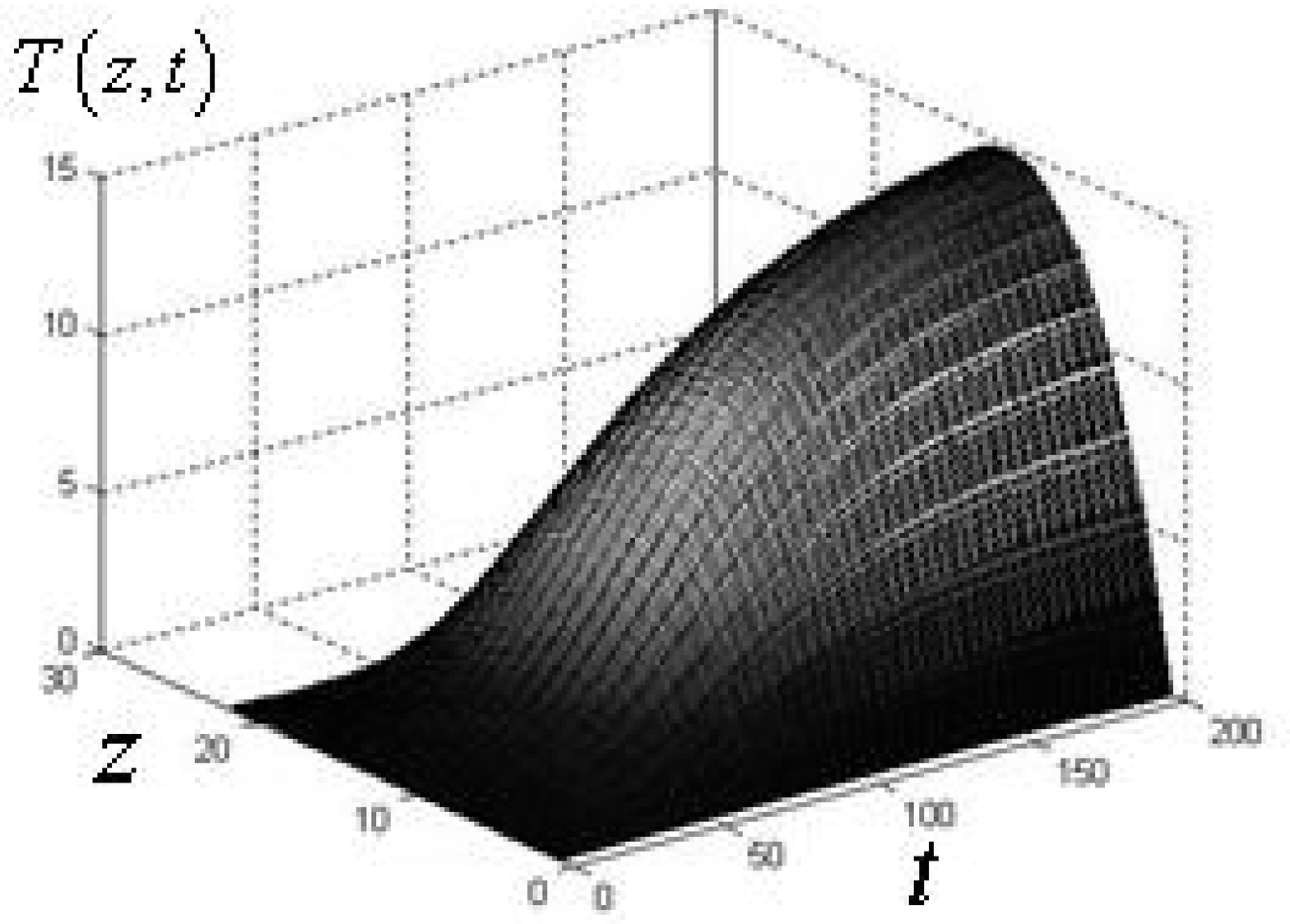}}}
\caption{System output.}
\end{figure}
\begin{figure}
\scalebox{0.25}[0.25]{\rotatebox{90} {\includegraphics{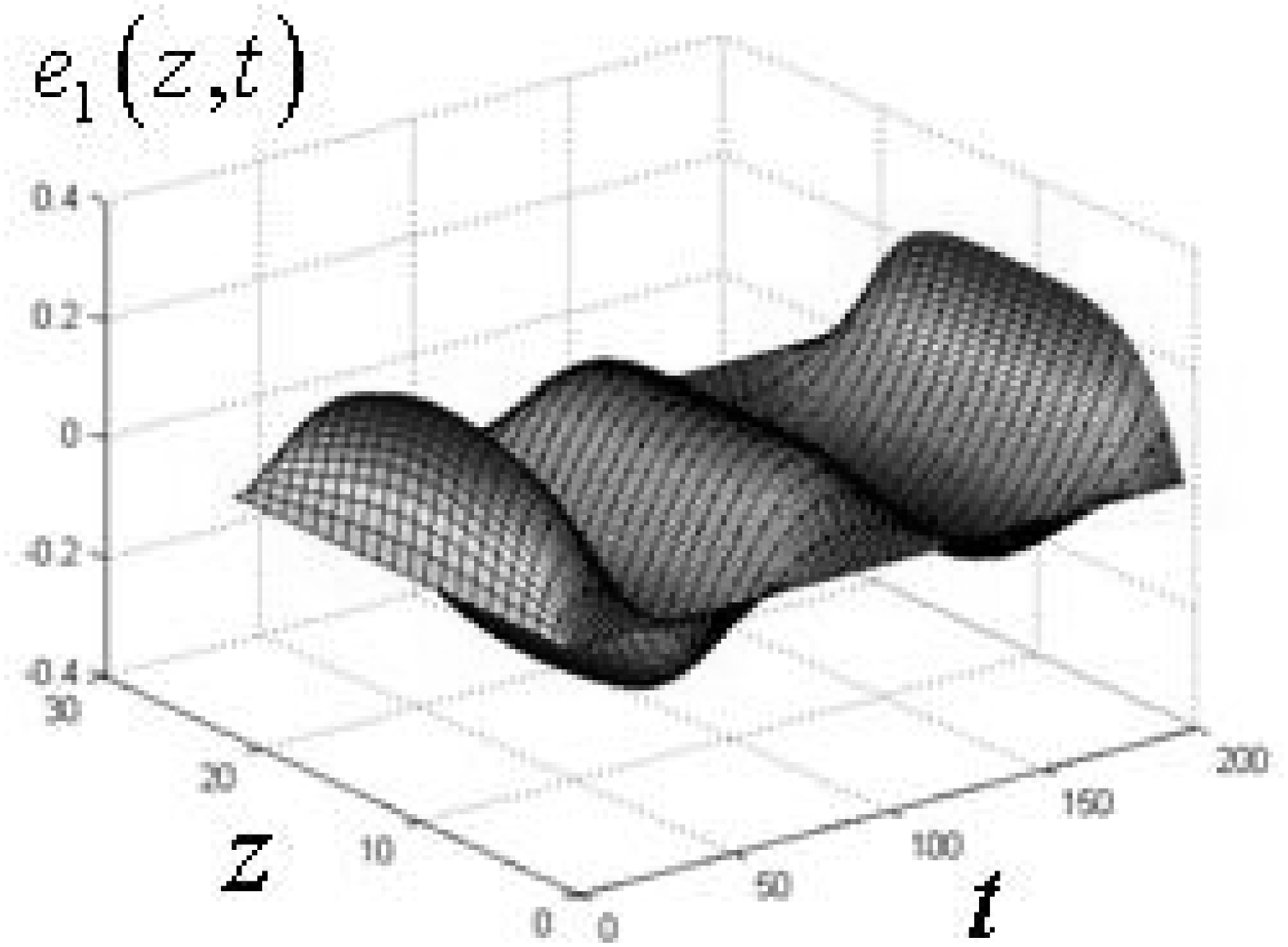}}}
\caption{Modeling error of the traditional method.}
\end{figure}
\begin{figure}
\scalebox{0.25}[0.25]{\rotatebox{90} {\includegraphics{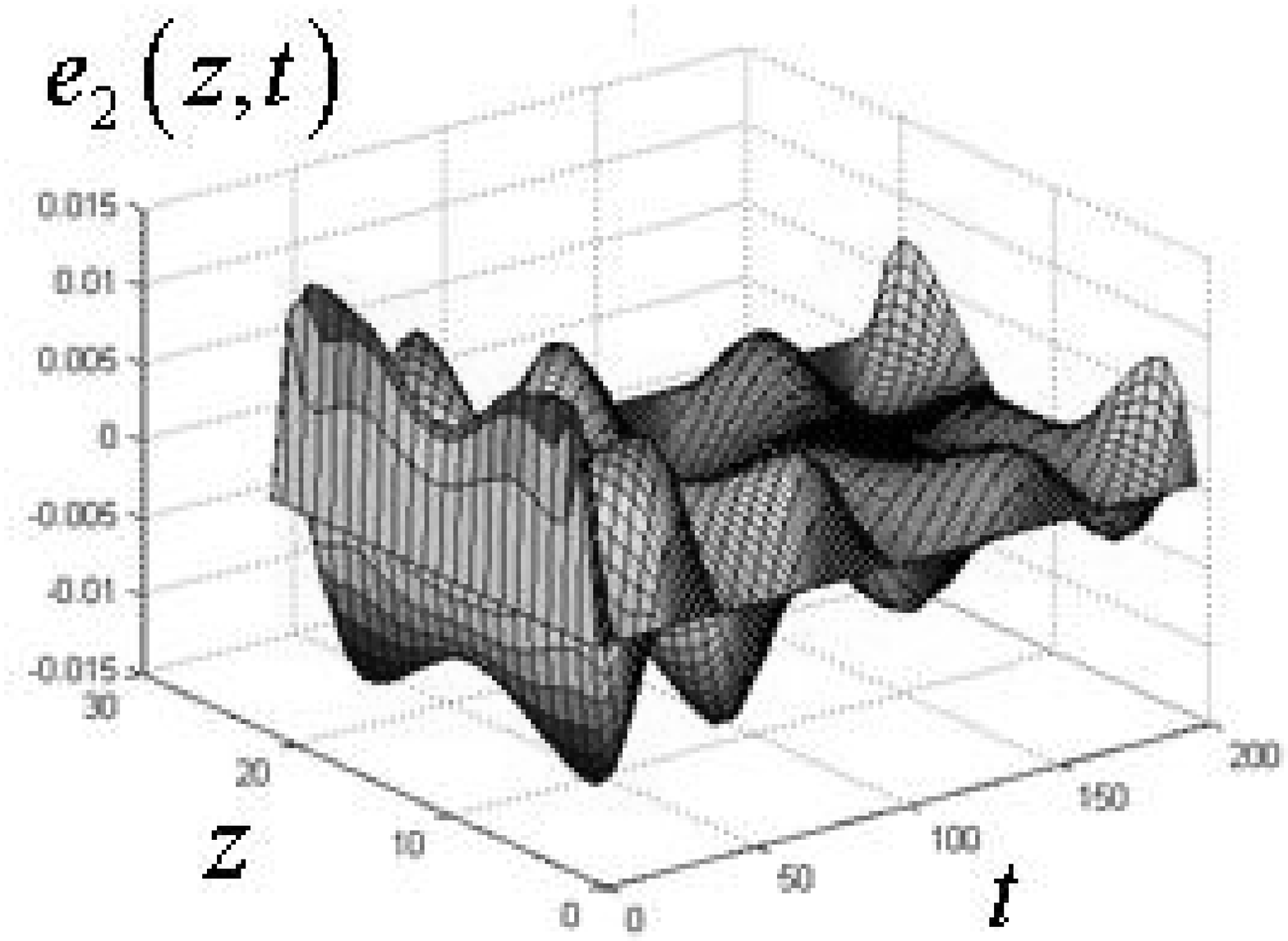}}}
\caption{Modeling error of the present method.}
\end{figure}

Fig. 2 shows the operation details of this modeling method. The
first order OFS filter is the Laguerre series, in which
\begin{equation}
G_0(q^{-1})=\frac{q^{-1}\sqrt{1-a^2}}{1-q^{-1}a},
G_1(q^{-1})=\frac{q^{-1}-a}{1-q^{-1}a},
\end{equation}
where $a$ is the time-scaling constant and $q^{-1}$ is the one
step backward shifting operator (i.e. $q^{-1}u(t)=u(t-1)$). The
second order OFS filter is the Kautz Series, in which
$G_0(q^{-1})$ and $G_1(q^{-1})$ are second order transfer
functions. Analogically, Heuberger \emph{et al.}
\cite{Heuberger1995} introduced the higher order OFS model. As the
order increases, OFS model can handle more complex dynamics.

\emph{Numerical Results} - Consider a long, thin rod in a reactor
as shown in Fig. 3. The reactor is fed with pure species $A$ and a
zeroth order exothermic catalytic reaction of the form
$A\rightarrow B$ takes place in the rod. Since the reaction is
exothermic, a cooling medium that is in contact with the rod is
used for cooling. Assume the density, heat capacity, conductivity
and temperature are all constant, and species $A$ is superfluous
in the furnace, then the mathematical model which describes the
spatiotemporal evolution of the rod temperature consists of the
following parabolic partial differential equation:
\begin{equation}
\frac{\partial T}{\partial t}=\frac{\partial^2T}{\partial
z^2}+\beta_Te^{-\frac{\gamma}{1+z}}-\beta_Te^{-\gamma}+\beta_{u}(b(z)u(t)-T),
\end{equation}
which subjects to the Dirichlet boundary conditions:
\begin{equation}
T(0,t)=0,\texttt{  }T(\pi,t)=0,\texttt{  }T(z,0)=0,
\end{equation}
where $T(z,t)$, $b(z)$, $\beta_T$, $\beta_{u}$, $\gamma$, and $u$
denote the temperature in the reactor (output), the actuator
distribution function, the heat of reaction, the heat transfer
coefficient, the activation energy, and the temperature of the
cooling medium (input), respectively. Here we set $\beta_T=50.0$,
$\beta_{u}=2.0$, and $\gamma=4.0$. In the numerical calculation,
without loss of generality, we set $b(z)=1$, and
$u(t)=[1.4,1.4,1.4,1.4]$. The order of Volterra series is two, the
OFS is chosen as one-order Laguerre series \cite{Datta1995} with
$a=0.6$, and the truncation length is given $N=4$.

\begin{figure}
\scalebox{0.22}[0.22]{\includegraphics{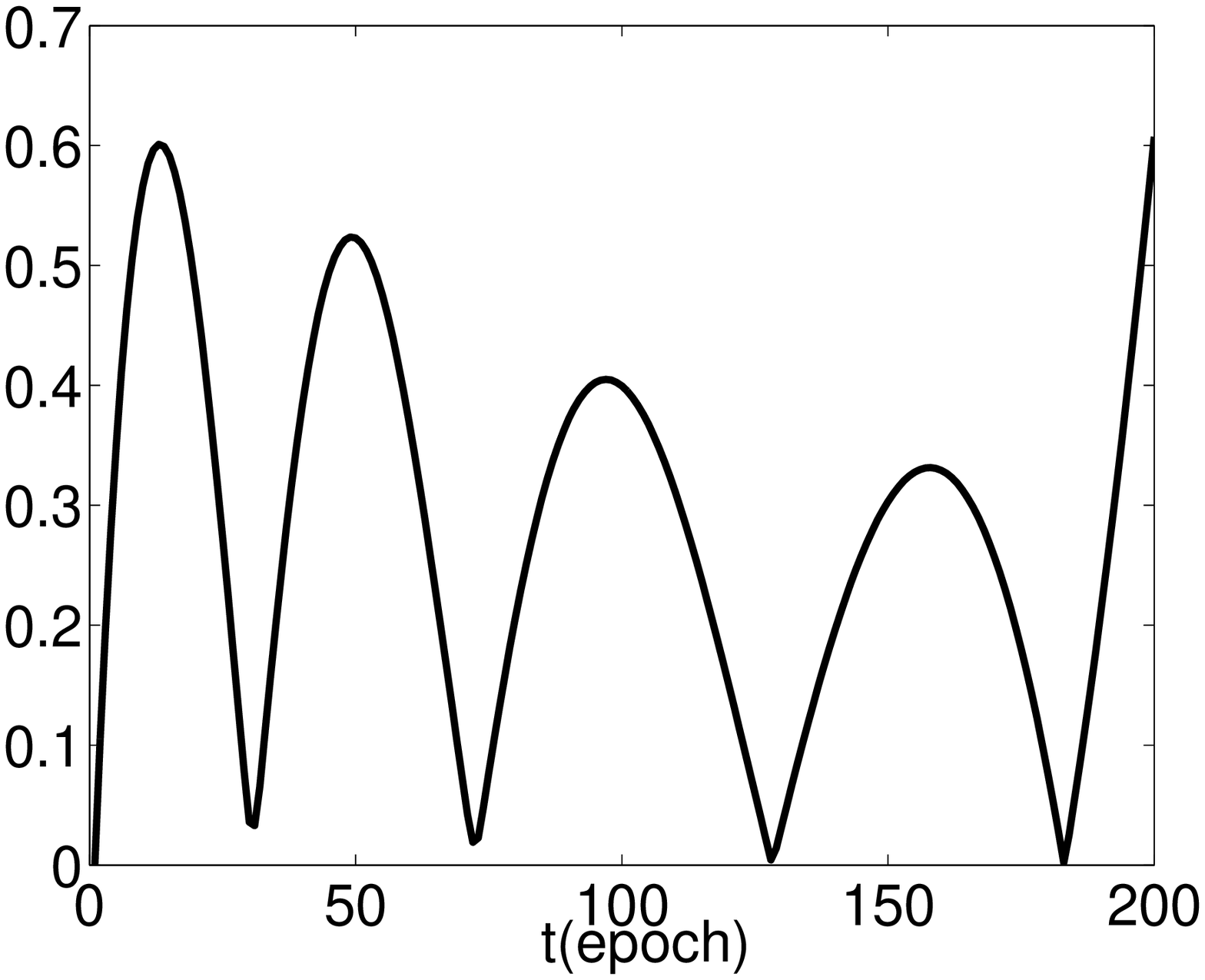}}
\scalebox{0.22}[0.22]{\includegraphics{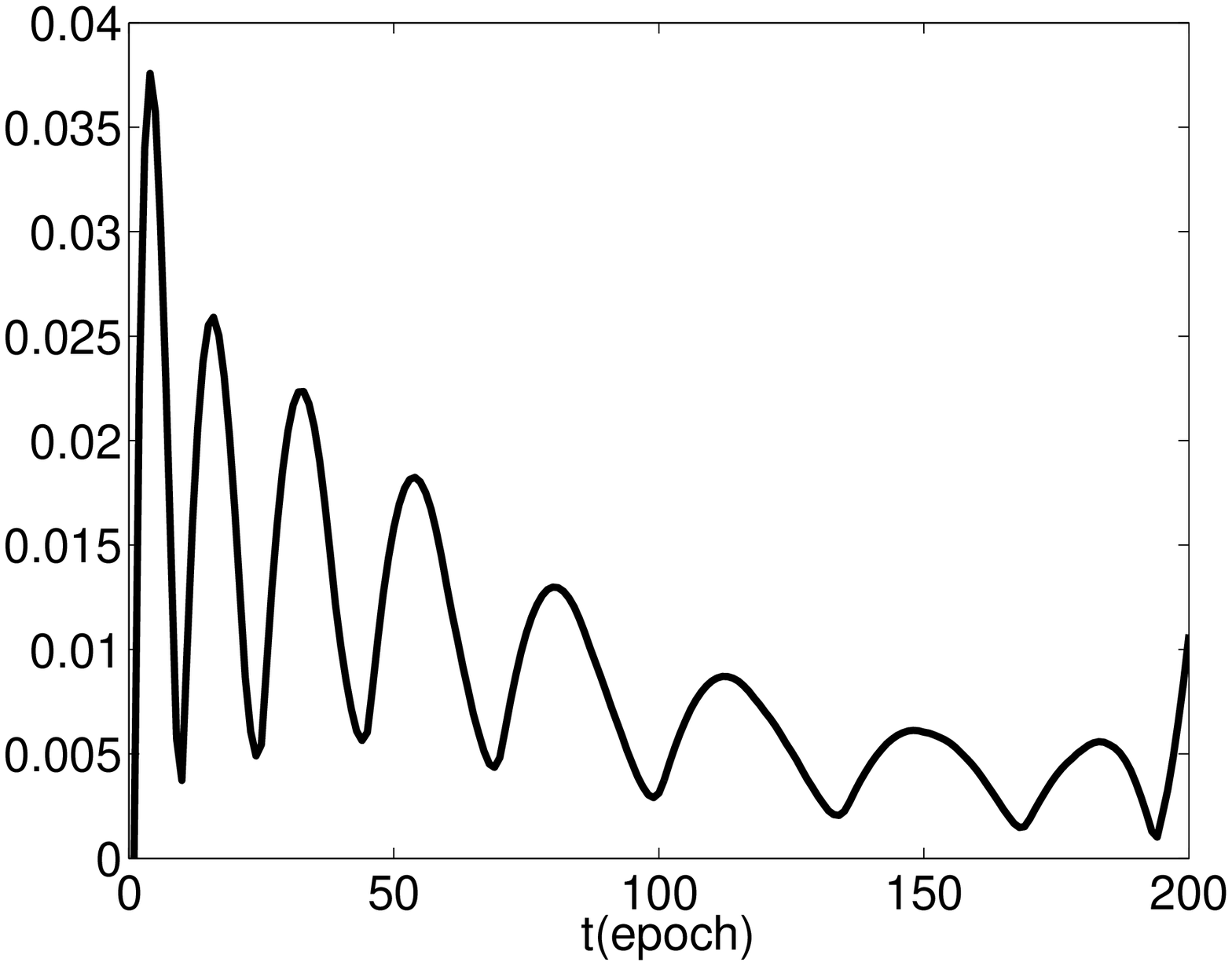}} \caption{IAEs of
the traditional method (left) and the present method (right).}
\end{figure}
\begin{figure}
\scalebox{0.22}[0.22]{\includegraphics{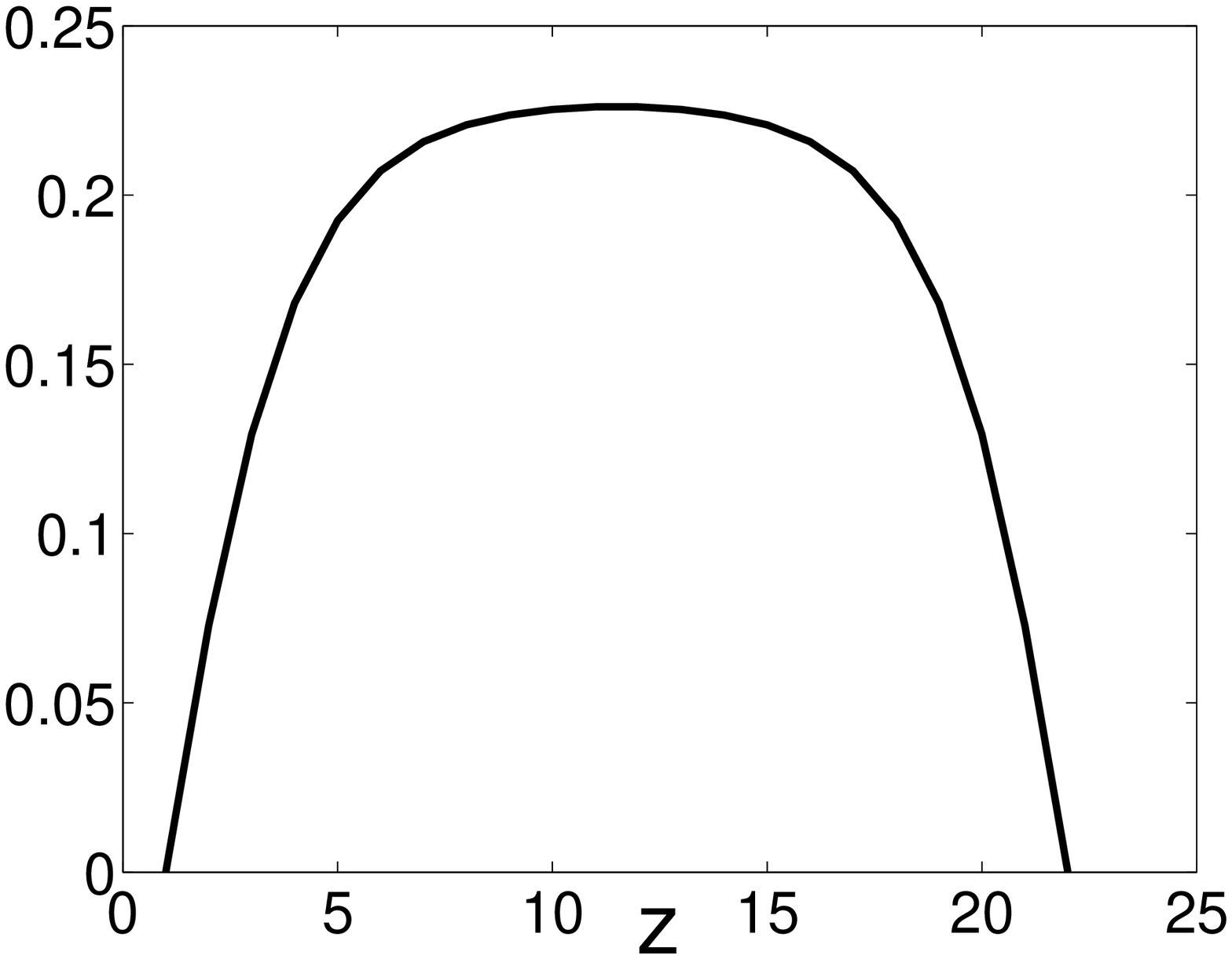}}
\scalebox{0.22}[0.22]{\includegraphics{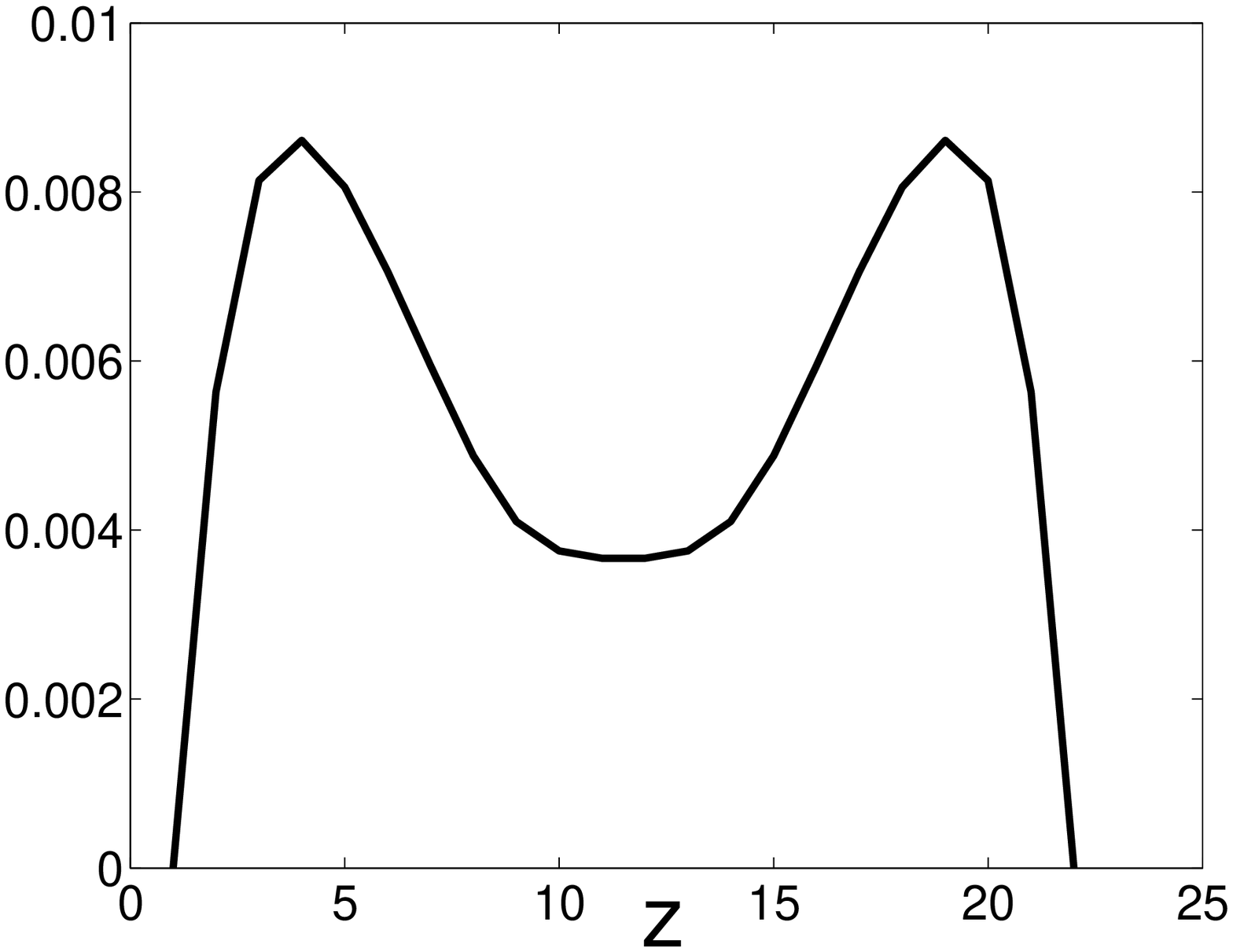}} \caption{ITAEs of
the traditional method (left) and the present method (right).}
\end{figure}
\begin{figure}
\scalebox{0.4}[0.5]{\includegraphics{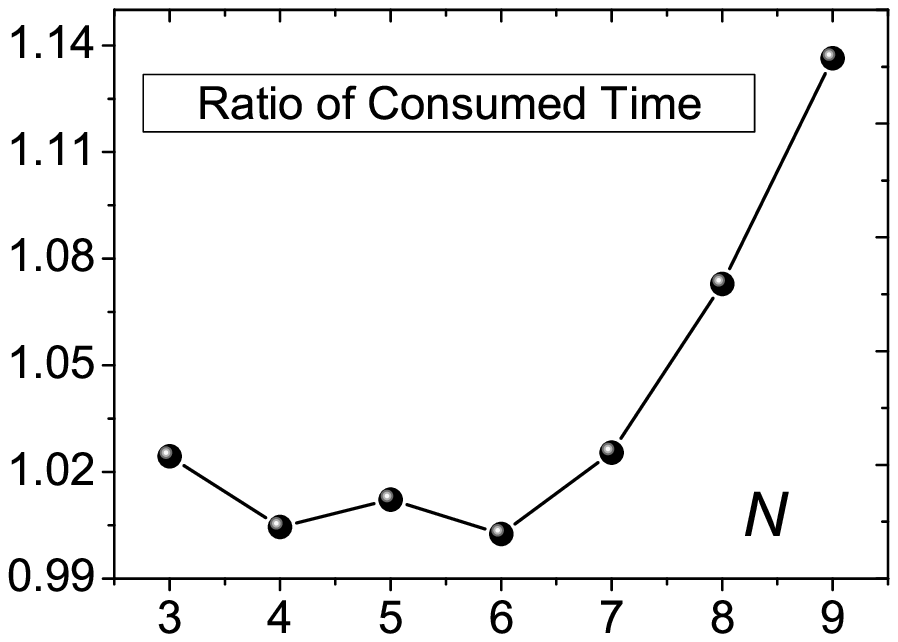}}
\scalebox{0.4}[0.5]{\includegraphics{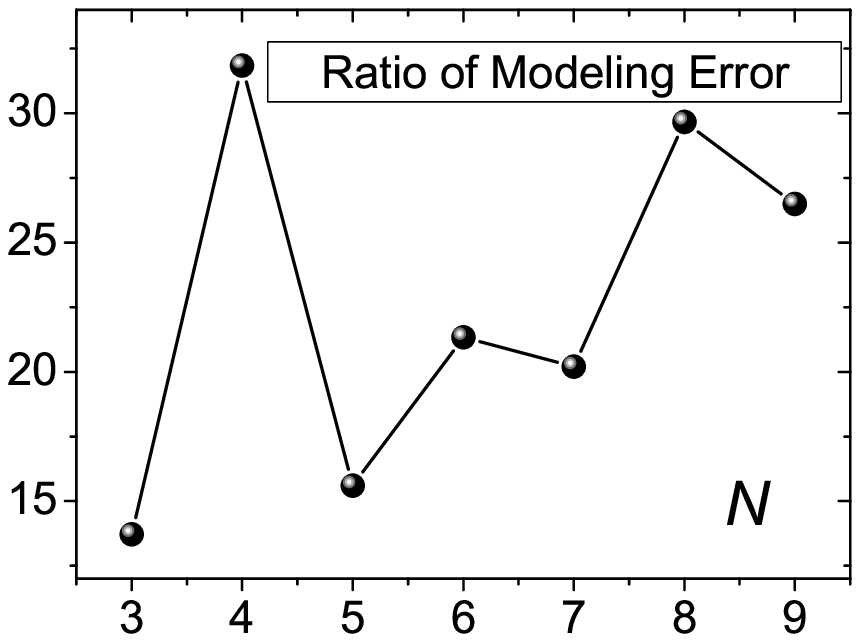}} \caption{Ratios of
the consumed time $t_2/t_1$ (left) and the average of absolute
modeling error $|e_1|/|e_2|$ (right) vs the truncation length $N$.
The subscripts 1 and 2 denote the cases of the traditional and the
present methods, respectively. The CPU time by using traditional
method for $N\in [3,9]$ is in the interval $[100s,160s]$. All the
numerical calculations are implemented by using a personal
computer with CPU: 1.8G, RAM: 256M, OS: Windows XP, and software
platform: MATLAB 6.5.}
\end{figure}

The system output is shown in Fig. 4. Denote by $e(z,t)$ the
modeling error, that is, the difference between system output and
modeling result at the point $(z,t)$. Fig. 5 and Fig. 6 exhibit
the modeling errors of the traditional and the present methods,
respectively. Clearly, the method proposed here has remarkably
smaller error than that of the traditional one. To provide a vivid
contrast between these two methods, we calculate the integral of
absolute error (IAE, $\int|e(z,t)|dz$) and time-weighted absolute
errors (ITAE, $\int t|e(z,t)|dt$), which are two standard error
indexes to evaluate modeling performances of DPS and can be
considered as the modeling errors along the temporal dimension $t$
and the spatial dimension $z$. As are shown in Fig. 7 and Fig. 8,
both the IAE and ITAE of the present method is reduced by $>20$
times comparing with those of the traditional one, which strongly
demonstrates the advantage of the present method. Actually, to
obtain the shapes of IAE and ITAE, one can cut the error surfaces
of Figs. 5 and 6 along the temporal coordinate $t$ and the spatial
dimension $z$. In addition, we calculate the average of absolute
modeling error $\frac{1}{\int\int dzdt}\int\int |e(z,t)|dzdt$.
From Fig. 9, it is found that, in comparison with the traditional
method, the modeling accuracy of the present one is enhanced by
14-32 times with less than $15\%$ increase of the consumed time.
It should be note that the modeling accuracy would increase along
with the increase of the Volterra series order $N_v$, however, the
computational complexity will increase too. For lumping systems,
this fact has been proved, and for DPSs, this fact is also
validated by experimental results \cite{Schetzen1980}. So a
tradeoff between the modeling accuracy and the computational
complexity must be made. That is why here we set the order
$N_v=2$.

\begin{figure}
\scalebox{0.25}[0.25]{\rotatebox{90} {\includegraphics{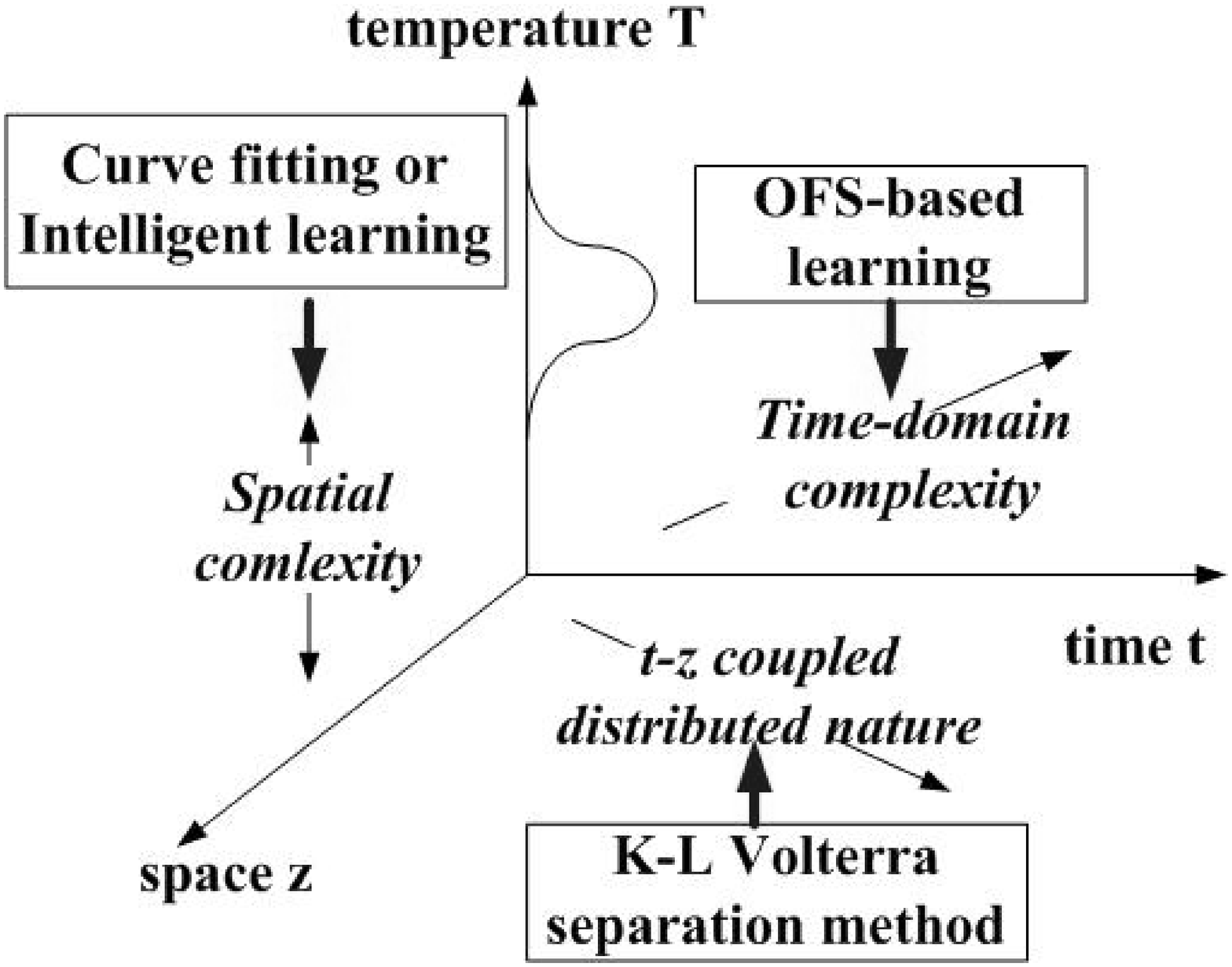}}}
\caption{Modeling methodology for DPSs.}
\end{figure}

\emph{Conclusion and Discussion} - Modeling method for nonlinear
DPS plays an important role in physical system analysis and
industrial engineering. Unfortunately, there exits two essential
difficulties in this issue, a) \textbf{distributed nature} due to
time-space coupled, which causes different temperature responses
at different locations; b) \textbf{nonlinear complexity} from
varying working point - different dynamics obtained even at the
same location for a large change of working points. Owing to these
difficulties, previous modeling methods via linear time/spatial
separation techniques (e.g. KL approach, spectrum analysis,
SVD-Galerkin technique, etc.) can not yield satisfying modeling
performance, especially for DPSs with severe nonlinearity. The
modeling error is caused by the nonlinear residue of the linear
time/space separation. Thus, it is naturally to expect that a
nonlinear time/space separation method may yield better modeling
performance.

To validate this supposition, we design a novel modeling method by
extending the concept of lumping Volterra series to the
distributed scenario. As shown in Fig. 10, the nonlinear DPS is
first decomposed into kernels, upon which the time-space
separation is carried out. These two decompositions will gradually
separate the complexity and provide a better modeling platform.
The time/space separation will be handled by a novel KL Volterra
method instead of the conventional KL method, the time-domain
complexity by the OFS-based learning, and the spatial complexity
by the curve fitting techniques (e.g. spline interpolation) or
intelligent learning algorithms (e.g. neural network, fuzzy
system, etc.). This novel method is applied on a benchmark
nonlinear DPS of industrial process, a catalytic rod. It is found
that the modeling accuracy is improved by more than 20 times in
average comparing with the traditional method, with almost no
additional computational complexity. The underlying reason may be
that the high order Volterra kernel can compensate the residuals
of the linear separation. In addition, we have applied this method
to another two benchmark nonlinear DPSs, a rapid thermal chemical
vapor deposition process \cite{Christofides2001}, and a
Czochralski crystal growth process \cite{Christofides2001}. The
corresponding results also strongly suggest that the nonlinear
time/space separation can greatly enhance the modeling accuracy.

Although its superority has been demonstrated, the KL Volterra
method is just a first attempt aiming at the motivation of
nonlinear time/space separation. Thanks to its excellent modeling
efficiency, this novel method is definitely a promising one for
both physical system analysis and industrial engineering. We
believe that our work can enlighten the readers on this
interesting subject.

The author would like to thank Prof. Guanrong Chen for helpful
discussion and suggestion. HTZhang would like to acknowledge the
National Natural Science Foundation of China (NSFC) under Grant
Nos. 60340420431 and 60274020, and the Youth Founding Project of
HUST. TZhou would like to acknowledge NSFC under Grant Nos.
70471033, 10635040 and 70571074. HXLi would like to acknowledge
the RGC-CERG founding of Hong Kong Government under Grant Nos.
9041015 and 9040916.

\end{document}